\documentclass[aps,pra,showpacs,showkeys]{revtex4}
\usepackage{epsfig}
\usepackage{hyperref}
\begin{document}

\def\il{I_{low}}
\def\iu{I_{up}} 
\def\eeq{\end{equation}} 
\def\ie{i.e.}  
\def\etal{{\itet al. }}  
\def\prb{Phys. Rev. {\bf B}} 
\def\pra{Phys. Rev. {\bf A}}
\def\prl{Phys. Rev. Lett. } 
\def\pla{Phys. Lett. A } 
\def\pb{PhysicaB} 
\def\pr{Phys. Rev.}  
\def\ajp{Am. J. Phys. }  
\def\jpc{J. Phys. C} 
\def\rmp{Rev. of Mod. Phys. }  
\def\jap{J. Appl. Phys. }
\def\mpl{Mod. Phys. Lett. {\bf B}} 
\def\ijmp{Int. J. Mod. Phys. {\bfB}} 
\def\ijp{Ind. J. Phys. } 
\def\ijpap{Ind. J. Pure Appl. Phys. }
\def\ibmjrd{IBM J. Res. Dev. } 
\def\pjp{Pramana J. Phys.}
\def\ssc{Solid State Comm. }

\title{Wave attenuation model for dephasing and measurement of
  conditional times}

\author{A. M. Jayannavar} 
\email{jayan@iopb.res.in}
\author{Colin Benjamin}
\email{colin@iopb.res.in}
\affiliation{Institute of Physics, Sachivalaya Marg, Bhubaneswar 751 005,
  Orissa, India}
\date{\today}

\begin{abstract}
  Inelastic scattering induces dephasing in mesoscopic systems. An
  analysis of previous models to simulate inelastic scattering in such
  systems is presented and also a relatively new model based on wave
  attenuation is introduced. The problem of Aharonov-Bohm(AB)
  oscillations in conductance of a mesoscopic ring is studied. We have
  shown that conductance is symmetric under flux reversal and
  visibility of AB oscillations decay to zero as function of the
  incoherence parameter, signalling dephasing.  Further wave
  attenuation is applied to a fundamental problem in quantum
  mechanics, i.e., the conditional(reflection/transmission) times
  spent in a given region of space by a quantum particle before
  scattering off from that region.
\end{abstract}
\maketitle

\section{Introduction}

We present a study of two different problems in mesoscopic physics
which are of current interest namely, dephasing and the other
conditional times. This work considers these two phenomena not from a
microscopic point of view but through a phenomenological model which
captures the essence of these quite well. This model is
known as wave attenuation. It essentially involves damping the wave
function in the region of interest, so as to derive essential physics.
In the context of dephasing, we consider an Aharonov-Bohm
interferometer, and show that the wave attenuation model is better
than its counterpart the optical potential model used in this context.
Also, in case of double heterostructures we employ this technique to
clock the time a particle takes to traverse a local region of interest
in the given system. Here also the wave attenuation model is more
simpler to deal with than other models.

\section{Dephasing}

The process of dephasing or decoherence leads to the diminishing of
quantum effects or loss of quantum mechanical interference effects.
Dephasing occurs due to interaction of an electron (interfering
entity) with it's environmental degrees of freedom (which are not
measured in the interference experiment)\cite{imry,fluorian}.  The AB
oscillations are one of the prime examples for analyzing how quantum
interference effects are affected by dephasing. These oscillations are
similar to the fringes seen in an Young's double slit experiment apart
from the presence of the magnetic flux. In the Young's double slit
interferometer the intensity is given by
$I=|\Psi|^2=|\psi_1|^2+|\psi_2|^2+2 \Re(\psi_1^*\psi_2^{} e^{i
  \phi})$, the part $2 \Re(\psi_1^*\psi_2^{} e^{i \phi})$ represents
the interference term. Here $\psi_1$ and $\psi_2$ are the complex wave
amplitudes across the upper and lower arms of the interferometer and
$\phi$ is phase difference between these two wave amplitudes.  If
there would have been no phase relationship between the waves then the
average intensity will be $<I>=|\psi_1|^2+|\psi_2|^2$. Complete
dephasing is indicated by extinction of these interference terms.
Thus, Dephasing can be defined as a phenomena by which quantum
mechanical system behave as though they are described by classical
probability theory.

\subsection{ Model's for dephasing} 

In absence of a complete microscopic theory as to how inelastic
scattering affects dephasing, models are useful.  There are different
ways to model dephasing in mesoscopic systems. An interesting method
is to attach a voltage probe\cite{buti} to the sample as in inset of
Figure~1 (Buttiker's model). In this model, an electron captured by a
voltage probe is re-injected back with an uncorrelated phase leading
to irreversible loss of phase memory. This model has built in current
conservation and Onsager's symmetry relations are obeyed but it does
suffer from a major demerit in that it only describes localized
dephasing.

The optical potential model provides another method of introducing
dephasing in these mesoscopic samples. The optical potential was first
introduced to explain the inelastic crossection in case neutron
scattering. In this model a spatially uniform optical potential($-i
V_i$) is added to the Hamiltonian\cite{zohta}, making it
non-hermitian.  This leads to removal of particles from phase coherent
motion, while absorption is real in the case of photons there is no
absorption for electrons.  The absorption of electrons is
reinterpreted\cite{yuming} as scattering into other energy levels and
therefore proper re-injection is necessary. Zohta and
Ezawa\cite{zohta} interpreted that the total transmission is defined
after re-injection as the sum of two contributions one due to the
coherent part and the other due to the incoherent part, i.e.,
$T_{tot}=T_{coh}+T_{incoh}$. The incoherent part is calculated as
$T_{incoh}=\frac{T_{r}}{T_{l}+T_{r}}A$, herein $T_{r}$ and $T_{l}$ are
the probabilities for right and left transmission from the region of
inelastic scattering and $A$ is the absorbed part which is given by
$A= 1-T_{coh}-R_{coh}$.  This model unlike Buttiker's model describes
dephasing that occurs throughout the system but still it has some
major drawbacks in that it violates Onsager's two terminal symmetry
relations\cite{pareek} and in the limit of strong absorption leads to
perfect reflection.
 
\begin{figure}[h]
\protect\centerline{\epsfxsize=3.0in \epsfbox{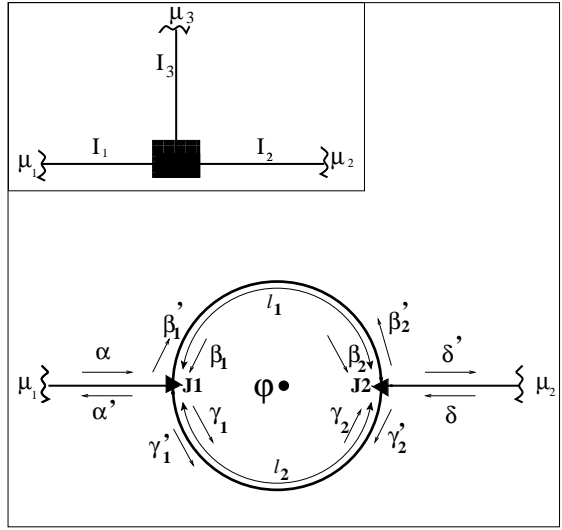}}
\caption{Aharonov-Bohm ring geometry. Inset shows the three probe model.}
\end{figure}

Thus, there is need of a model which is free from the shortcomings of
the above two models. Brouwer and Beenakker\cite{brouwer} have
developed a simple method, by mapping the three probe Buttiker's
method into a two terminal geometry, this is done by eliminating the
transmission coefficients which explicitly depend on the third probe
by means of unitarity of the S-matrix.  They consider a three
terminal geometry in which one of the probes is used as a voltage
probe in absence of magnetic flux (see inset of Figure~1).  A current
$I=I_1=-I_2$ flows from source to drain. In this model, a fictitious
third lead connects the ring to a reservoir at chemical potential
$\mu_3$ in such a way that no current is drawn $(I_3=0)$.  The $3$X$3$
S-matrix of the entire system can be written as-

\[S=\left(\begin{array}{ccc}
r_{11}& t_{12}&t_{13}\\
t_{21}& r_{22}&t_{23}\\
t_{31}& t_{32}&r_{33}
\end{array} \right) \]

Application of the relations\cite{buti,brouwer,datta}- $I_p=\sum_q
G_{pq}[\mu_p-\mu_q], p=1,2,3$ and $ G_{pq}=(2e^2/h)T_{pq}$ yields the
(dimensionless) two probe conductance
$G=\frac{h}{2e^2}\frac{I}{\mu_1-\mu_2}$,

\begin{eqnarray}
G=T_{21}+\frac{T_{23}T_{31}}{T_{31}+T_{32}}
\end{eqnarray}

where $T_{pq}=T_{p\leftarrow q}$, is the transmission from $q^{th}$ to
$p^{th}$ lead ($|t_{pq}|^2$), on elimination of the transmission
coefficients in Eq.~1, which involve the voltage probe using the
unitarity of the S - Matrix leads \cite{brouwer} to-

\begin{eqnarray}
G=T_{21}+\frac{(1-R_{11}-T_{21})(1-R_{22}-T_{21})}{1-R_{11}-T_{21}+1-R_{22}-T_{12}}.
\end{eqnarray}

Now all the above coefficients are built from the 2X2 S matrix-
\[S^{\prime}=\left(\begin{array}{cc}
r_{11}& t_{12}\\
t_{21}& r_{22}\\
\end{array} \right) \]
which represents the S Matrix of the absorbing system
(non-hermitian)\cite{CB}. Thus re-injection has been reformulated as
in Eq.~2. The first term in Eq.~2 represents the conductance
contribution from the phase coherent part.  The second term accounts
for electrons that are re-injected from the phase breaking reservoir,
thereby ensuring particle conservation in the voltage probe model.
Also, Eq.~2 restores Onsager's two terminal symmetry relation for the
optical potential model.

\subsection{Wave attenuation as a model for dephasing}

In-spite of the fact that a major problem associated with the optical
potential has been cured there still remains the problem that in the
strong absorption limit it leads to perfect reflection and absorption
without reflection(spurious scattering) is not possible
\cite{joshi,jayan,rubio}. To overcome this problem, we instead of
making the Hamiltonian non-hermitian, add an exponential factor
$e^{-\alpha l}$ to the S-Matrix of the system.  This is the model of
wave attenuation. This describes dephasing which occurs throughout the
system and removes the short comings of the optical potential model.

The wave attenuation model is not new it has earlier been dealt with
in the context of 1-D localization\cite{joshi}. In the AB ring
geometry considered here, wave attenuation is inserted by the factor
$e^{-\alpha l_1}$ (or $e^{-\alpha l_2}$) in the complex free
propagator amplitudes, every time we traverse\cite{datta} the upper
(or lower) arms of the ring (see Figure~1). We have calculated the
relevant transmission and reflection coefficients by using the S-
matrix method along with the quantum wave guide theory for a single
channel case. In this model, average absorption per unit length is
given by $2\alpha$.  With this method we show that the calculated
conductance $(G)$ in Eq.~2 is symmetric under the flux reversal as
required. The visibility of the AB oscillations rapidly decay as a
function of $\alpha$, indicating dephasing. Hence forth we refer to
$\alpha$ as an incoherence parameter. Increasing $\alpha$ corresponds
to increasing dephasing processes in the system or increase in
temperature.

In Figure~1, the length of the upper arm is $l_1$ and that of lower
arm is $l_2$. The total circumference of the loop is $L=l_1+l_2$. The
loop is connected to two current leads.  The couplers (triangles) in
Figure~1 which connect the leads and the loop are described by a
scattering matrix $S$. The S matrix for the left coupler yields the
amplitudes $O_{1}=(\alpha^\prime,\beta_{1}^\prime,\gamma_{1}^\prime)$
emanating from the coupler in terms of the incident waves
$I_1=(\alpha,\beta_{1},\gamma_{1})$, and for the right coupler yields
the amplitudes
$O_{2}=(\delta^\prime,\beta_{2}^\prime,\gamma_{2}^\prime)$ emanating
from the coupler in terms of the incident waves
$I_2=(\delta,\beta_{2},\gamma_{2})$. The S-matrix for either of the
couplers\cite{butipra} is given by-

\[S=\left(\begin{array}{ccc}
-(a+b)       & \sqrt\epsilon&\sqrt\epsilon\\
\sqrt\epsilon& a            &b            \\
\sqrt\epsilon& b            &a            
\end{array} \right) \]

with $a=\frac{1}{2}(\sqrt{(1-2\epsilon)} -1)$ and
$b=\frac{1}{2}(\sqrt{(1-2\epsilon)} +1)$. $\epsilon$ plays the role of
a coupling parameter. The maximum coupling between reservoir and loop
is $\epsilon=\frac{1}{2}$, and for $\epsilon=0$, the coupler
completely disconnects the loop from the reservoir.

The waves incident into the branches of the loop are related by the S
Matrices for upper branch by-

\[\left(\begin{array}{c}
\beta_1\\
\beta_2\\
\end{array} \right) \ =\left(\begin{array}{cc}
0     & e^{ikl_1} e^{-\alpha l_1} e^\frac{-i \theta l_1}{L}\\
e^{ikl_1} e^{-\alpha l_1} e^\frac{i \theta l_1}{L} & 0 \\
\end{array} \right) \left(\begin{array}{c}
\beta_1^\prime\\
\beta_2^\prime
\end{array} \right)\]  
and  for lower branch-

\[\left(\begin{array}{c}
\gamma_1\\
\gamma_2\\
\end{array} \right) \ =\left(\begin{array}{cc}
0     & e^{ikl_2} e^{-\alpha l_2} e^\frac{i \theta l_2}{L}\\
e^{ikl_2} e^{-\alpha l_2} e^\frac{-i \theta l_2}{L} & 0 \\
\end{array} \right) \left(\begin{array}{c}
\gamma_1^\prime\\
\gamma_2^\prime
\end{array} \right)\]

These S matrices of course are not unitary
$S(\alpha)S(\alpha)^\dagger\neq 1 $ but they obey the relation $
S(\alpha)S(-\alpha)^\dagger= 1 $. The same relation is also obeyed by
the S Matrix of the system in presence of imaginary potential.  Here
$kl_1$ and $kl_2$ are the phase increments of the wave function in
absence of flux.  $\frac{\theta l_1}{L}$ and $\frac{\theta l_2}{L}$
are the phase shifts due to flux in the upper and lower branches.
Clearly, $\frac{\theta l_1}{L}+\frac{\theta
  l_2}{L}=\frac{2\pi\Phi}{\Phi_0} $, where $\Phi$ is the flux piercing
the loop and $\Phi_0$ is the flux quantum$\frac{hc}{e}$. The
transmission and reflection coefficients in Eq.~2 are given as
follows- $T_{21}=|\frac{\delta^\prime}{\alpha}|^2$,
$R_{11}=|\frac{\alpha^\prime}{\alpha}|^2$,
$R_{22}=|\frac{\delta^\prime}{\delta}|^2$,
$T_{12}=|\frac{\alpha^\prime}{\delta}|^2$ wherein
$\delta^\prime,\delta,\alpha^\prime,\alpha$ are as depicted in
Figure~1.

After calculating the required reflection and transmission
coefficients we see that the coherent transmission $T_{21}$ is not
symmetric under the flux reversal however proper re-injection of
carriers by Eq.~2 for the total conductance $G$ plotted shows that the
Onsager's symmetry relations are restored, i.e., $G$ is symmetric
under flux reversal. On can notice from this figure that amplitude of
AB Oscillations decrease with increase in incoherence parameter
$\alpha$. All parameters used in the following figures are in their
dimensionless form.
\begin{figure} [h]
\protect\centerline{\epsfxsize=3.5in\epsfbox{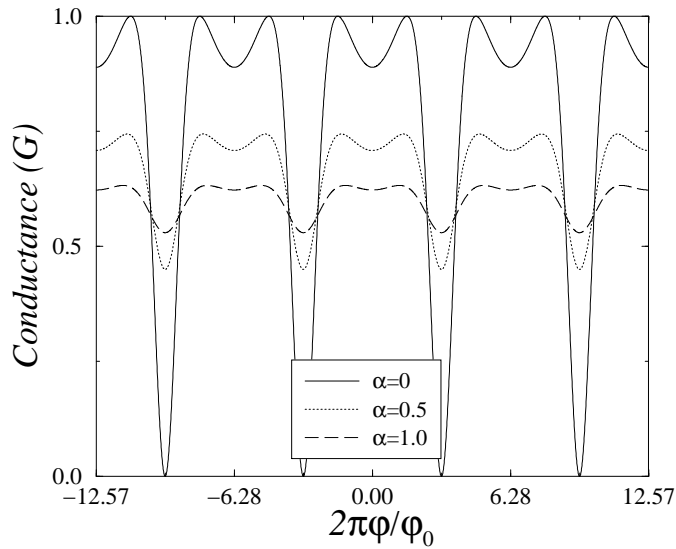}}
\caption{Reduction of Aharonov-Bohm Oscillations in presence of
  incoherence parameter $\alpha$. $kL=\pi$, $l_{1}/L=0.75$, $l_{2}/L=0.25$, and
  coupling $\epsilon=0.5$. }
\end{figure}

In Fig.~3 we plot visibility ($V$) as a function of incoherence parameter
$\alpha$. Visibility is of course defined as-
\begin{eqnarray}
V=\frac{G_{max}-G_{min}}{G_{max}+G_{min}}.
\end{eqnarray}

\begin{figure} [h]
\protect\centerline{\epsfxsize=3.5in\epsfbox{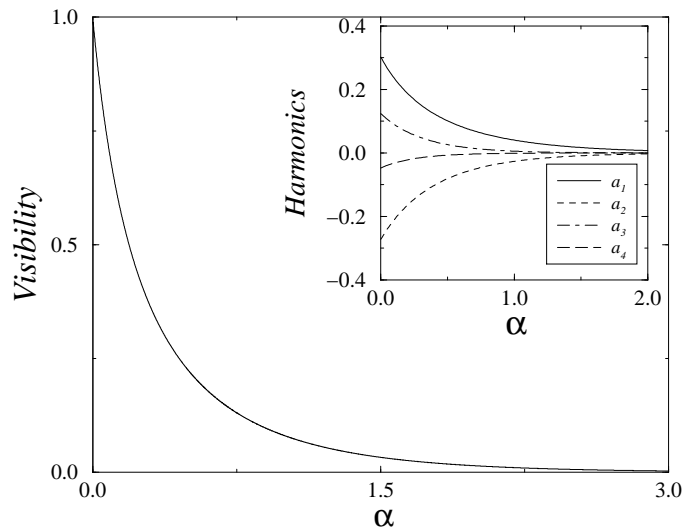}}
\caption{Visibility
 for the same physical parameters as in Figure~2, coupling parameter
 $\epsilon=0.5$. In the inset the harmonics have
 been plotted for the same physical and coupling parameters.}
\end{figure}

The plot shows that with increase in the value of the parameter
$\alpha$ the visibility exponentially falls off, reaching a point
where it becomes zero corresponding to the disappearance of quantum
interference effects, i.e., total dephasing in the system. In the
inset of Figure~3 we have plotted the first few Fourier\cite{xia}
harmonics $a_n$ (wherein $n=1$ to $4$) of $G(\Phi)$ as a function of
$\alpha$.  The harmonics are calculated from the
formula-$a_{n}=\int_{0}^{\pi}G cos(n\phi)d\phi$.  The harmonics
exponentially fall off with increasing $\alpha$ with exponent
increasing as we go from $1$ to $4$. The $n^{th}$ order harmonic
corresponds to the contribution from electronic paths which encircle
the flux $n$ times.  The harmonics can sometimes show non monotonic
behavior depending on the physical parameters\cite{joshids}, however,
the visibility is a monotonic function of the incoherence parameter
$\alpha$.

Thus we have seen that wave attenuation can be effective in modeling
absorption induced dephasing in mesoscopic systems. In the next section 
we will see how this method is extended to measure the time a quantum
particle takes to traverse a specified region of scattering.

\section{Time in quantum mechanics}

One of the most important problem in quantum mechanics is to calculate
the time spent by a particle in a given region of space before
scattering off from that region. The problem is essentially due to the
fact that there is no hermitian operator to calculate this time in
quantum mechanics\cite{landauer,hauge,gasparian}. The prospect of
nanoscale electronic devices has in recent years brought new urgency
to this problem as this is directly related to the maximum attainable
speed of such devices.  When it comes to tunneling time or time in
general there is lot of ongoing controversy. In some formulations this
time leads to a real quantity and in others to a complex
quantity\cite{baskin}. In certain cases tunneling time is considered
to be ill-defined or quantum mechanics does not allow us to discuss
this time\cite{baskin,yamada,steinberg}. Furthermore sometimes it is
maintained that tunneling through a barrier takes zero
time\cite{reflandauernature}. Recently, Anantha Ramakrishna and Kumar
(AK)\cite{anantha} have proposed the non unitary Optical potential as
a clock to calculate the sojourn times without the clock affecting it.
In this paper we examine another non-unitary clock, i.e.,wave
attenuation to calculate the conditional lengths, i.e., the total
effective distance traveled by a particle in the region of interest.
This conditional length on appropriate division by the speed of the
particle in the region of interest will give us the conditional time.

The criteria, any result for the time spent by a quantum particle in a
given region of scattering should satisfy are that-(1) It should be
real, (2) It should add up for non overlapping regions, (3) It should
be causally related to the interval of space, and (4) tend to the
correct classical limits. It is shown explicitly by AK that all clock
mechanisms involve spurious scattering or very clock mechanism affects
the sojourn times to be clocked finitely even as the perturbation due
to clock potential is infinitesimally small. All of these clocks
involve spurious scattering as the perturbation due to clock mechanism
couples to the Hamiltonian. This raises the question ``Can quantum
mechanical sojourn time be clocked with clock affecting it?''. In this
paper we introduce such a method in which perturbation is not
introduced in the Hamiltonian but in the S-Matrix of the system. In
this case scattering is treated analogously with the Fabry-Perot
interferometer. The scheme is illustrated in Figure~4.

\subsection{Wave attenuation  to measure conditional times}  

In the wave attenuation method\cite{datta,joshi}, we damp the wave
function by adding an exponential factor ($e^{-\alpha l}$) every time
we traverse the length of interest, here $2\alpha$ represents the
attenuation per unit length. This method is better than the optical
potential model as it does not suffer from spurious
scattering's\cite{CB,jayan,rubio}. The corrections introduced in case
of Optical potential model to take care of spurious scattering's will
become manifestly difficult when we calculate the traversal times for a
superlattice involving numerous scatterer's. Thus our method of wave
attenuation scores over the optical potential model. In the presence
of wave attenuation a wave attenuates exponentially and thus the
transmission (or reflection) coefficient becomes exponential with the
length endured in presence of the attenuator and this acts as a
natural counter for the sojourn lengths. Following the procedure of AK
we calculate the traversal and reflection lengths and times in a given 
region of interest (in particular between two scatterer's as in Figure~4.

\begin{figure}[h]

\protect\centerline{\epsfxsize=4.0in \epsfbox{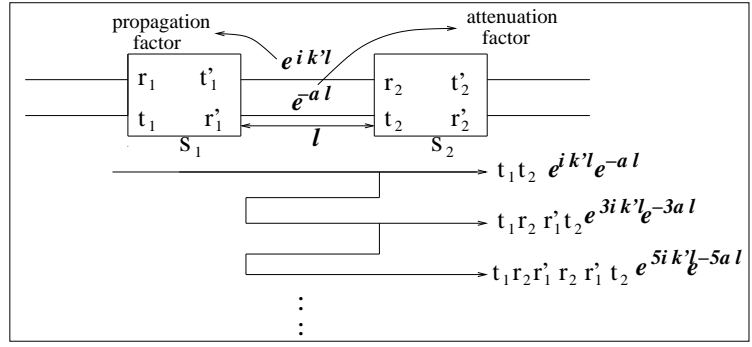}}
\caption{Summing the different paths, $S_1$ and $S_2$ denote the two
  scatterer's. $l$ is the distance between them. $e^{i
    k^{\prime}l}$ and $e^{-\alpha l}$ denote the propagation and attenuation
    factors in the locality of interest.}
\end{figure}

The amplitude for transmission and reflection can be
calculated by summing\cite{datta} the different paths as in Figure~4.
The scatterer $S_{1}$ in Figure~1 has as its elements
$r_{1}^{},r_{1}^{\prime},t_1^{}$ and $t_1^{\prime}$. $r_{1}^{}$ is the
reflection amplitude when a particle is reflected from the left side
of the barrier while $r_{1}^{\prime}$ is the reflection amplitude when
a particle is reflected from the right side of the barrier. $t_1^{}$
and $t_1^{\prime}$ are the amplitudes for transmission when a particle
is transmitted from left to right of the barrier and vice-versa.
Similar assignments are done for the scatterer $S_2^{}$.

Thus for the amplitude of transmission we have-
$t=t_{1}^{}t_{2}^{}e^{i k^{\prime}l}e^{-\alpha l}
+t_{1}^{}r_{2}^{}r^{\prime}_{1}t_{2}^{}e^{3 i k^{\prime}l}e^{-3 \alpha
  l} +...$ which can be summed as
\begin{eqnarray}
t=\frac{t_{1}^{}t_{2^{}}e^{i k^{\prime}l}e^{-\alpha
           l}}{1-r_{2}^{}r^{\prime}_{1}e^{2 i k^{\prime}l}e^{-2 \alpha l} }
\end{eqnarray}
and this is the transmission amplitude in presence of wave
attenuation.  Again for the case of reflection amplitude we have
$r=r_{1}^{}+t_{1}^{}r_{2}^{}t^{\prime}_{1}e^{2i k^{\prime}l}e^{-2\alpha
  l}+t_{1}^{}r_{2}^{}r^{\prime}_{1}r_{2}^{}t^{\prime}_{1}e^{4 i
  k^{\prime}l}e^{-4 \alpha l}+
t_{1}^{}r_{2}^{}r^{\prime}_{1}r_{2}^{}r^{\prime}_{1}r_{2}^{}t^{\prime}_{1}e^{6
  i k^{\prime}l}e^{-6 \alpha l}+.. $ , which leads to -

\begin{eqnarray}
r=\frac{r_{1}^{}-a r_{2}^{}e^{2i k^{\prime}l}e^{-2\alpha
           l}}{1-r^{\prime}_{1}r_{2}^{}e^{2 i k^{\prime}l}e^{-2 \alpha l} }
\end{eqnarray}

In Eq.~5, $a=r_{1}^{}r^{\prime}_{1}-t_{1}^{}t^{\prime}_{1}$ is the
determinant of the S-Matrix of the first barrier and as we are only
dealing with unitary S-Matrices therefore the determinant is of
unit modulus for all barriers.  In these expressions
$k^{\prime}$ is the wave vector in the region of interest.  The
transmission and reflection coefficients can be calculated by taking
the square of the modulus of the expressions in Eq's.~(4) and (5).

The traversal length for transmission and reflection length in case of 
reflection are calculated as below\cite{anantha}-

\begin{eqnarray}
l^{T}=\lim_{2\alpha \rightarrow 0} -\frac{\partial (\ln
  |t|^2)}{\partial (2 \alpha)} 
\end{eqnarray}
and the reflection length in case of reflection is defined as- 
\begin{eqnarray}
l^{R}=\lim_{2\alpha \rightarrow 0} - \frac{\partial (\ln
  |r|^2)}{\partial (2 \alpha)} 
\end{eqnarray}

The traversal times for transmission or reflection times in case of
reflection can be calculated from the formula-
$\tau^{R/T}=\frac{l^{R/T}}{\frac{\hbar k^{\prime}}{m}}$ wherein as
before $\frac{\hbar k^{\prime}}{m}$ is the speed of propagation in the
region of interest. For the case of the potential profile sketched in
Figure~5, $k^{\prime}=k$. From Eq's.~(6) and (7) we can calculate the
traversal length for transmission-
\begin{eqnarray}
\frac{l^{T}}{l}=\frac{1-|r^{\prime}_{1}|^{2}|r_{2}^{}|^{2}}{1-2
  \Re(r^{\prime}_{1}r_{2}^{} e^{2 i k^{\prime} l}) +|r^{\prime}_{1}|^{2}|r_{2}^{}|^{2}}
\end{eqnarray}
and for reflection-
\begin{eqnarray}
\frac{l^{R}}{l}=\frac{l^{T}}{l}+\frac{|r_{2}^{}|^{2}-|r_{1}^{}|^{2}}{|r_{1}^{}|^{2}-2
  \Re(r^{*}_{1}r_{2}^{}a  e^{2 i k^{\prime} l}) +|r_{2}^{}|^{2}}
\end{eqnarray}
Here $\Re$ represents real part of the quantity in brackets.  In the
above two equations the traversal and reflection lengths have been
normalized with respect to the length $l$ of the locality of interest,
which is the well region of the potential profile of Figure~5. Throughout
the discussion the quantities are expressed in their dimensionless
form.

\begin{figure}[h]
\protect\centerline{\epsfxsize=4.0in \epsfbox{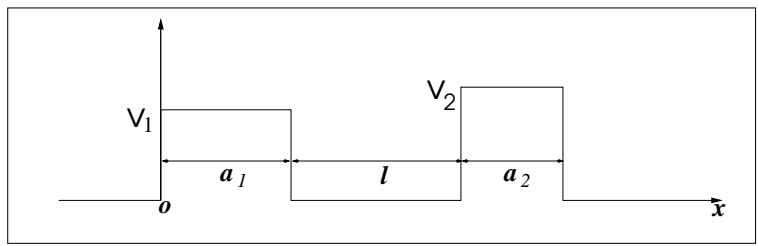}}
\caption{The double heterostructure. Barrier's are denoted by length
  $a_{j}$'s, height $V_j$'s and $j=1,2$. The length of the well is $l$.}
\end{figure}

We consider the case of two rectangular barriers  as
shown in Figure~5 separated by a distance $l$. The energies, potentials
and the lengths are all in their dimensionless form.

The S-Matrices for the double heterostructure are given as-
\[S_j=\left(\begin{array}{cc}
    r_j^{} & t_j^{\prime}\\
    t_j^{} & r_j^{\prime}\\
\end{array} \right) \ =\left(\begin{array}{cc}
\frac{-i\epsilon_{j+}\sinh K_{j}a_{j}}{2\cosh
  K_{j}a_{j}-i\epsilon_{j-}\sinh K_{j}a_{j}}     & \frac{2e^{-i k a_j}}{2\cosh
  K_{j}a_{j}-i\epsilon_{j-}\sinh K_{j}a_{j}} \\
\frac{2 e^{-i k a_j}}{2\cosh K_{j}a_{j}-i\epsilon_{j-}\sinh K_{j}a_{j}}     &
\frac{-e^{-2ika_{j}} i \epsilon_{j+} \sinh K_{j}a_{j}}{2\cosh K_{j}a_{j}-i\epsilon_{j-}\sinh K_{j}a_{j}} \\
\end{array} \right)\]  

wherein $K_{j}=\sqrt{\frac{2m(V_{j}-E)}{\hbar^2}}$,
$k=\sqrt{\frac{2mE}{\hbar^2}}$and
$\epsilon_{j\pm}=\frac{k}{K_j}\pm\frac{Kj}{k}$. Herein, $j$=1,2.

\begin{figure}[h]
\protect\centerline{\epsfxsize=3.5in \epsfbox{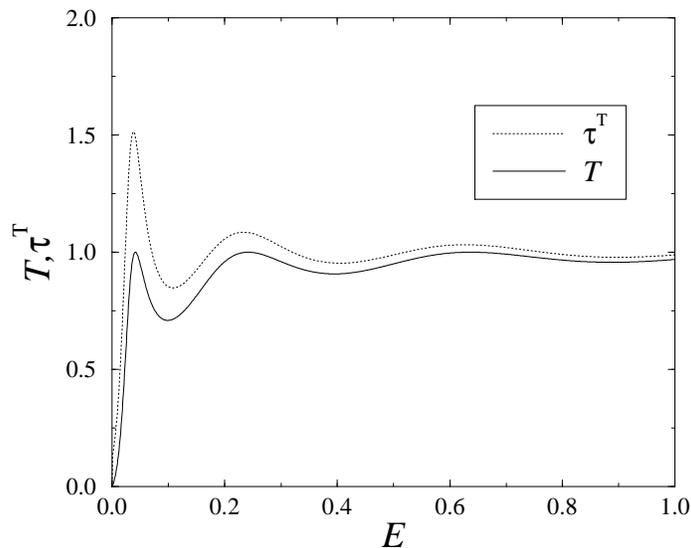}}
\caption{$T$ and $\tau^{T}$ for a  symmetric double heterostructure
  . $l=10.0$, $V_{1}=V_{2}=1.0$ and $a_{1}=a_{2}=0.2$.}
\end{figure}

In Figure~6 we plot the Transmission coefficient and the normalized
times $\tau^{T}$ (=$\tau^{R})$ for a symmetrical double barrier
$V=V_{1}=V_{2}$. These times are those required by a quantum particle
to traverse or reflect from the well of the double heterostructure. We
have normalized these times by the time taken by a particle to
traverse the length $l$, i.e., $\frac{m l}{\hbar k^{\prime}}$ without
the potential profile as in Figure~5. From Eq.~8, it is clear that
transmission time is always positive. Moreover, we can readily show
from our treatment that local times spent by the particle in
non-overlapping regions are additive.  As expected the traversal times
are larger near the resonances. In case of non-symmetrical structures
one obtains negative reflection times (which should not be surprising)
in some parameter region but traversal times are positive for both
symmetrical as well as non-symmetrical structures. It can be
argued\cite{anantha} that this is because in case of reflection there
is a partial wave corresponding to prompt reflection $r_{1}$ that
never samples the region of interest, and also this prompt part leads
to self interference delays which cause the sojourn time $\tau^{R}$ to
become negative for some values of the parameters. If one removes this
prompt part, i.e., $r_{np}=r-r_{1}^{}$, and we re-calculate the
sojourn time $\tau^{R_{np}}$ with this prompt part removed we find it
to be positive and given by $\tau^{R_{np}}=\tau^{T}+1$, as $\tau^{T}$
is positive. In the case of non-symmetrical structures $\tau^{T}$ is
independent of the fact that a particle is incident from the left or
the right, but $\tau^{R}$ depends on the direction of the incident
particle. There is also a remarkable assertion found in the
literature\cite{hauge} concerning the measurement of the time of
transmission or reflection, which is $\tau^{D}=T\tau^{T}+R\tau^{R}$.
Herein $\tau^{T}$ and $\tau^{R}$ are as given above while the dwell
time $\tau^{D}=\frac{1}{v}\int_{0}^{l} |\psi|^2 dx$. $\psi$ is the
wavefunction in the locality of interest and $v$ is the speed of the
particle in the region of interest.  We have verified explicity after
calculation that this equivalence does not hold \cite{landauer,CBssc}.
Our method can also be readily extended to calculation of traversal
time of tunneling (for case of a single barrier). Results are in
agreement with recent calculations\cite{anantha}.

\section {Conclusion}

In conclusion, we have shown that wave attenuation is much better in
modeling dephasing due to absorption than it's counterpart the optical
potential. Further when we extend this method to calculate the
conditional sojourn times we find here also wave attenuation is easier
to deal with than optical potential.


\begin{thebibliography} {99}
  
\bibitem{imry} Y.Imry, {\it Introduction to Mesoscopic Physics}
  (Oxford University Press, New York, 1997).
 
\bibitem{fluorian} Florian Marquardt, {\it An Introduction to the
    basics of dephasing}
  (http://iff.physik.unibas.ch/~florian/dephasing/dephasing.html).
 
\bibitem{buti} M. Buttiker, \prb {\bf 33}, 3020 (1986);\ibmjrd {\bf
    32}, 63 (1988).
  
\bibitem{zohta} Y. Zohta and H. Ezawa, \jap {\bf 72}, 3584 (1992).

\bibitem{yuming} Hu Yuming, \jpc {\bf 21}, L23 (1988).
\bibitem{pareek} T. P. Pareek, Sandeep K. Joshi, and A. M. Jayannavar, \prb
   {\bf 57}, 8809 (1998);  

\bibitem{brouwer} P. W. Brouwer and C. W. J. Beenakker, \prb {\bf
    55}, 4695 (1997); P. W. Brouwer, Ph.D. thesis, Insttuut-Lorentz,
  University of Leiden, The Netherlands,1997.

\bibitem{datta} S. Datta, {\it Electron Transport in mesoscopic
    systems} (Cambridge University press, Cambridge, 1995).
  
\bibitem{CB} Colin Benjamin and A. M. Jayannavar, \prb {\bf 65},
  153309 (2002); preprint cond-mat/0112153.

\bibitem{joshi} Sandeep K. Joshi, D. Sahoo and A. M. Jayannavar, \prb
   {\bf 62}, 880 (2000); P. Pradhan, preprint cond-mat/9807312.

\bibitem{jayan} A. M. Jayannavar, \prb {\bf 49}, 14718 (1994); A. K.
  Gupta and A. M. Jayannavar,\prb {\bf 52}, 4156 (1995).
  
\bibitem{rubio} A. Rubio and N. Kumar, \prb {\bf 47}, 2420 (1993).

\bibitem{butipra} M. Buttiker, Y. Imry and M. Ya. Azbel, \pra {\bf
      30}, 1982 (1984).

\bibitem{xia}
J. B. Xia, \prb {\bf 45}, 3593 (1992).

\bibitem{joshids} S. K. Joshi, D. Sahoo and A. M. Jayannavar, \prb {\bf
    64}, 075320 (2001).

\bibitem{landauer}
R. Landauer and Th. Martin, \rmp {\bf 66}, 217 (1994).

\bibitem{hauge} 
E. Hauge  and J. A. Stovneng, \rmp {\bf 61}, 917 (1989).

\bibitem{gasparian} V . Gasparian, M. Ortuno, G. Schon and U. Simon,
  {\it Tunneling time in Nanostructures}
  (http://bohr.fcu.um.es/papers/1999/gasparian.pdf).

\bibitem{baskin} 
D. Sokolovski and L. M. Baskin, \pra {\bf 36}, 4604 (1987).

\bibitem{yamada} 
N. Yamada, \prl {\bf 83}, 3350 (1999).

\bibitem{steinberg} 
A. Steinberg, \prl {\bf 74}, 2405 (1995).

\bibitem{reflandauernature} E. O. Kane, in {\it Tunneling Phenomena in
    Solids,} edited by E. Burstein and S. Lundquist (Plenum, New
  York), p.1 (1969).
 
\bibitem{anantha} S. Anantha Ramakrishna and N. Kumar, preprint
  cond-mat/0009269; S. Anantha Ramakrishna, Ph. D Thesis, Raman
  Research Institute, Bangalore, India (2001).
  
\bibitem{CBssc} Colin Benjamin and A. M. Jayannavar, \ssc {\bf 121}, 591 (2002); preprint cond-mat/0112499.

\end{thebibliography}
\end{document}